\documentstyle[12pt]{article}
\newlength{\dinwidth}
\newlength{\dinmargin}
\newcommand{\ba}{\begin{array}}
\newcommand{\ea}{\end{array}}
\newcommand{\be}{\begin{equation}}
\newcommand{\ee}{\end{equation}}
\newcommand{\bea}{\begin{eqnarray}}
\newcommand{\eea}{\end{eqnarray}}

\def\bee{\begin{eqnarray}}
\def\eee{\end{eqnarray}}
\def\be{\begin{equation}}
\def\ee{\end{equation}}

\newcommand{\beas}{\begin{eqnarray*}}
\newcommand{\eeas}{\end{eqnarray*}}

\font\cmss = cmss12

\def\integer{{\rlap{\cmss Z} \hskip 1.8pt \hbox{\cmss Z}}}
\def\laplace{{\kern1pt\vbox{\hrule height 1.2pt\hbox{\vrule width 1.2pt\hskip
  3pt\vbox{\vskip 6pt}\hskip 3pt\vrule width 0.6pt}\hrule height 0.6pt}
  \kern1pt}}
\def\scriptlap{{\kern1pt\vbox{\hrule height 0.8pt\hbox{\vrule width 0.8pt
  \hskip2pt\vbox{\vskip 4pt}\hskip 2pt\vrule width 0.4pt}\hrule height 0.4pt}
  \kern1pt}}

\def\roughly#1{\raise.3ex\hbox{$#1$\kern-.75em\lower1ex\hbox{$\sim$}}}

\begin{document}
\thispagestyle{empty}
\addtocounter{page}{-1}
\begin{flushright}
SNUTP 97-154\\
{\tt hep-th/9712055}\\
\end{flushright}
\vspace*{1.3cm}
\centerline{\Large \bf M(atrix) Theory on the Negative Light-Front
\footnote{
Work supported in part by the NSF-KOSEF Bilateral Grant, 
KOSEF SRC-Program, Ministry of Education Grant BSRI 97-2418 and The
Korea Foundation for Advanced Studies Faculty Fellowship}}
\vspace*{1.2cm} \centerline{\large\bf Soo-Jong Rey}
\vspace*{0.8cm}
\centerline{\large\it Physics Department, Seoul National University,
Seoul 151-742 KOREA}
\vspace*{0.6cm}
\centerline{\large\tt sjrey@gravity.snu.ac.kr}
\vspace*{1.5cm}
\centerline{\large\bf abstract}
\vskip0.5cm
M(atrix) theory defines light-front description of M-theory boosted along
positive direction of eleventh, M-coordinate. Rank of M(atrix) gauge group
is directly related to M-momentum $P_{11} = {\tt N} / R_{11}$ or, equivalently,
to total number of D0-partons. Alternatively, M-theory may be boosted along
opposite direction of M-coordinate, for which the theory consists only of
anti-D0 partons. In M(atrix) theory description, we interpret this as analytic
continuation of dimension of the gauge group: $\tt U(-N) \approx U(N)$, $\tt
SO(-2N) \approx USp(2N)$ and $\tt USp(-2N) \approx SO(2N)$. We check these 
`reciprocity relations' 
explicitly for uncompactified, heterotic, and CHL M(atrix) theories as well as
effective M(atrix) gauge theories of ${\bf T}_5/\integer_2$ and ${\bf T}_9/
\integer_2$ compactifications.
In all cases, we show that absence of parity, gauge and supersymmetry anomalies
require introduction of a twisted sector with negative numbers of matter
multiplets. They are interpreted as massless open string excitations connected
to anti-D-brane background.

\vspace*{1.1cm}

\setlength{\baselineskip}{18pt}
\setlength{\parskip}{12pt}

\newpage
{\tt Motivation:}
M(atrix) theory~\cite{bfss} defines the light-front Hamiltonian description 
of M-theory. By boosting Type IIA string theory along quantum, M-direction, 
parton content of M-theory is identified with D0-branes. Kinematical argument 
that leads to the identification of D0-partons is as follows.
Consider, in M-theory, excitation of a particle of rest mass $m$ and
eleven-momentum $(P_0, {\bf P}_\perp, P_{11})$:
\be
P_0^2 - P_{11}^2 - {\bf P}_\perp^2 = m^2.
\ee
If the particle is boosted by $P_{11} = N / R$ along ${\bf S}_1$ compactified
M-coordinate of radius $R$ while keeping ${\bf P}_\perp$ finite,
\bee
P_0 - |P_{11}|  &=& {1 \over P_0 + |P_{11}|}
\left( {\bf P}_\perp^2 + m^2 \right)
\nonumber \\
&\rightarrow& {1 \over 2 |P_{11}|} \left( {\bf P}_\perp^2 + m^2 \right)
\nonumber \\
&\equiv& {\cal H}_{\rm LC}
\eee
is of order ${\cal O}( 1/|P_{11}| )$. Along globally well-defined light-front
time $t \equiv X^0 + X^{11}$, ${\cal H}_{\rm LC}$ serves as the light-front 
Hamiltonian. The D0-particle corresponds to M-theory graviton ($m=0$) 
propagating around ${\bf S}_1$-compactified M-coordinate. Via Kaluza-Klein
interpretation, boost momentum $N/R_{11}$ is then identified with total
energy of D0-partons. Since $1/R_{11}$ is the minimal unit of the boost 
momentum, we identify $N$ with the total number of D0-particles. BPS nature
of Kaluza-Klein states imply that there are no contributions of 
anti-D0-particles. In other words, parton content of M(atrix) theory is 
entirely of D0-branes but no anti-D0-branes. This is the basis that 
M(atrix) theory is described by large-$N$ limit of ${\cal N} = 16$, 
$(0+1)$-dimensional $U(N)$ gauge theory. It is also possible that finite-$N$ 
M(atrix) theory is the discrete light-cone quantization (DLCQ) descriptio of 
M-theory~\cite{susskind}. Again, $N$ is directly related to the dimension of 
Fock space of D0-particles.

Given that, in M(atrix) theory, it appears that there is no room for 
anti-D0-particles. However, a variant of M(atrix) theory can be found, for 
which fundamental partons are made solely of anti-D0-branes. 
The idea is extremely simple that one boosts the M-theory backward (rather
than forward) along M-direction.
Magnitude of total M-momentum increases indefintely but the sign is {\sl 
opposite} to that of traditional forward boost.
For ${\bf S}_1$-compactified M-coordinates, this means that $N$ is taken a 
large negative integer: $P_{11} = N/R_{11} \rightarrow  - \infty$.
The Kaluza-Klein picture of the D0-particle Ramond-Ramond charge and CPT 
invariance tells us that boost with the negative $N$ should be interpreted as 
depleting $|N|$ D0-particles from Type IIA string, hence, attaching $|N|$
anti-D0-particles to it. 

That the sign of $P_{11}$ is correlated with particle or anti-particle 
interpretation can be seen more explicitly. Take infinite momentum boost
of a bosonic particle or anti-particle.  
In terms of light-front variables, the propagator is expressed as:
\bee
K(p) &\equiv& {i \over P^2 - m^2 \pm i \epsilon} 
\nonumber \\
&\rightarrow& {i \over E - {\cal H}_{\rm LC} + i \epsilon \, {\rm sgn} P_{11} }
\, \, .
\eee
The positive respectively negative $P_{11}$ momentum is then identified with 
particle and anti-particle excitations in the light-front description.

The above arguments suggest that, in M(atrix) theory, M-theory boosted 
backward along M-direction is described by $(0+1)$-dimensional gauge theory 
in which $N$ is replaced by $-N$ ($N > 0)$.
From Eq.~(1)
one finds that the change is to take the dimension of gauge group to a 
negative value!
How can one make sense of matrices whose sizes are negative-dimensional?
This paper is intended to show that it is indeed possible to give 
well-defined notion of M(atrix) theories with anti-D0-partons via so-called
`reciprocity relations' in the representation theory of Lie groups. 

{\tt Negative-Dimensional M(atrix) Gauge Groups} 
In different contexts in mathematics~\cite{penrose,king} and 
physics~\cite{cvitanovic,dunne}, classical groups with
negative dimensions have been considered over the years.
In particular, Cvitanovic and Kennedy~\cite{cvitanovic} have proven important \sl reciprocity
relations \rm, according which (1)  
for all $3m-j$ coefficients constructed from tensor representations of $\tt
SU(N)$, the interchange of symmetric and anti-symmetric Young tableaux is 
equivalent to the analytic continuation $\tt N \rightarrow - N$, and 
(2) for all $3m-j$ coefficients constructed from tensor representations
of $\tt SO(2N)$ and $\tt USp(2N)$, the interchange of symmetric and 
anti-symmetric
Young tableaux is equivalent to the analytic continuation $\tt N \rightarrow
- N$ together with the interchange $\tt SO \leftrightarrow USp$.
The reciprocity relations may be expressed symbolically as 
\bee
{\tt SU(-N)} &:=: & [{\tt SU(N)}]^{\rm T}
\nonumber \\
{\tt SO(-2N)} &:=:& [{\tt USp(2N)}]^{\rm T}
\nonumber \\
{\tt USp(-2N)} &:=:& [{\tt SO(2N)}]^{\rm T} \, ,
\eee
where ${}^{\rm T}$ signfies weighted transpose of Young tableaux, in which
the weighting is given by $(-)^m$ for $m$-boxed Young tableaux.  

In a way reminiscent to the idea of Parisi and Sourlas~\cite{parisisourlas}, 
one may interpret representations of these negative-dimensional classical 
groups as Grassmannian representations~\cite{dunne}.
Note that the analytic continuation $\tt N \rightarrow - N$ makes a sense
for any physical gauge invariant quanties. Even though there are in principle
ambiguities in changing sign of $\tt N$ for quantities that are defined only 
for integer positive $N$, in perturbation theory, the group weight of each 
Feynman diagram may be expressed in terms of eigenvalues of the Casimir 
operators on irreducible representations, hence, is a polynomial in $\tt N$. 
It is precisely for this polynomial type that they make sense for 
{\sl all} values of $\tt N$ via analytic continuation.
We now apply the above reciprocity relations to several known M(atrix) 
theories and check various consistency conditions.

{\tt M(atrix) Theory with Sixteen Supercharges:}
Let us begin with uncompactified M(atrix) theory~\cite{bfss}. 
The M(atrix) theory is described by $(0+1)$-dimensional
gauge theory (quantum mechanics) with sixteen supercharges and $\tt U(N)$
gauge group. The field content is gauge multiplet $(A_0)$ and a matter
multiplet $({\bf X}^i, {\bf \Theta}_\alpha)$ transforming in the adjoint
representation. As such, the reciprocity relation Eq.~(4) implies that all 
physical quantities calculated out of M(atrix) theory via perturbation theory
are identical in one-to-one manner with those calculated out of M(atrix) 
theory with gauge group $SU(-N)$. The latter is what we have identified with 
M(atrix) theory of anti-D0-partons. This should be intuitively clear. 
The M-theory is invariant under simultaneous reversal transformation of parity 
$X^{11} \rightarrow - X^{11}$ and three-form gauge potential 
$C_{MNP} \rightarrow - C_{MNP}$.
Reversing the boost direction amounts to Kaluza-Klein charge-conjugation of 
Type IIA string theory, hence, interchange of D0-partons and anti-D0-partons.
The fact that the three-form potential reverses the sign can be seen from, 
for example, the {\sl sign} of non-commutativity of the membrane configuration 
$[{\bf X}^i, {\bf X}^j] = \epsilon^{ij} / N$.
Upon ${\bf T}^n$ toroidal compactification, the M(atrix) theory is described by 
maximally supersymmetric gauge theories with gauge group $\tt U(-N) 
:=: [U(N)]^{\rm T}$. The $(n+1)$-dimensional gauge theory is vector-like, 
hence, is free from potential parity, gauge or supersymmetry anomalies.
We thus conclude that M(atrix) theory with negative-dimensional gauge group
$\tt U(-N)$ corresponds to the light-front description of M-theory boosted 
backward in M-direction and the negative-dimension of the gauge group is 
interpreted as parton content being anti-D0-particles rather than D0-particles.

{\tt Heterotic M(atrix) Theory:}
Heterotic M(atrix) theory is defined as ${\bf S}_1/\integer_2$ orbifold of
M(atrix) theory and is described by $(0+1)$-dimensional gauge theory
with eight supercharges and $\tt O(2N)$ gauge group.
An important distinction of heterotic M(atrix) theory is that the theory
requires introduction of a twisted sector. 
The field content of untwisted sector consists of gauge multiplet $(A_0)$, 
ajoint multiplet $(A_9, {\bf S}_a)$, and a matter multiplet 
$({\bf X}^i, {\bf S}_{\dot a})$ that transform as rank-two symmetric 
representation under the gauge group. In Ref.~\cite{kimrey1}, it has been 
shown that local cancellation of cosmological constant require introduction 
of a twisted sector consisting of sixteen, fermionic supermultiplets 
$\chi_M^{(1,2)}$ that transform under $\tt O(2N)$ gauge group as fundamental
representations.
In the limit of shrinking ${\bf S}_1/\integer_2$ orbifold, 
the heterotic M(atrix) theory becomes $(1+1)$-dimensional {\sl chiral} gauge
theory with $(8, 0)$ supersymmetry and gauge group $\tt O(2N)$~\cite{rey}. 
The gauge and 
adjoint multiplets of quantum mechanics combine into $(1+1)$-dimensional gauge 
supermultiplet $(A_\mu, {\bf S}_a)$. 
Being chiral, the gauge theory is severely constrained by absence of potential 
gauge and supersymmetry anomalies~\cite{kimrey1, kabatrey}.  
The gauge anomaly of untwisted sector $8 \, C_2 ({\tt adjoint}) - 
8 \, C_2 ({\tt symmetric}) = 8 [ ({\tt N}  + 2) - ({\tt N} - 2) ] 
C_2 ({\tt fundamental})$
is cancelled precisely by the gauge anomaly 
$ {\tt N_F} C_2 ({\tt fundamental})$ 
of the twisted sector spectrum, viz. $\tt N_F = 32$ Majorana-Weyl 
fermion supermultiplets transforming in the fundamental 
representations.

Consider now heterotic M(atrix) theory with anti-D0-partons. Reversal of
boost momentum is achieved by analytic continuation of $\tt N \rightarrow
- N$, hence, M(atrix) gauge group $\tt SO(-2N)$. 
According to the `reciprocity relations', this has to be accompanied
by the weighted transpose of Young tableaux for each multiplets. 
The gauge multiplet,
which corresponds to anti-symmetric rank-two representation, is turned to
symmetric rank-two representation. This implies that $\tt SO(-2N) = USp(2N)$
and is a confirmation of the `reciprocity relation'.    
At the same time, the multiplicity of twisted sector multiplets has to be
taken negative value $\tt N_F \rightarrow - N_F$.
In weakly coupled Type-I string limit, the twisted sector spectrum is 
understood as being massless chiral fermionic excitations of D1-D9 strings.
Now that the M-theory boost direction is reversed, hence, D1 and D9 branes
are turned into anti-D1 and D9 branes, the anti-D1-anti-D9 string gives rise 
to {\sl negative} number of chiral fermions. 
They are anti-particles of chiral fermions measured relative to the M(atrix)
theory vacuum.
In other words, $\tt N_F \rightarrow - N_F$ is a reflection of the fact that
the background D-branes are turned into anti-D-branes.

We thus conclude that heterotic M(atrix) theory obtained by backward boost
is described by a gauge theory whose gauge group is $\tt SO(-2N) = USp(2N)$ 
and multiplicity of twisted sector spectrum is taken to be negative, 
$\tt N_F = - 32$ in units of fundamental representations. 

{\tt CHL M(atrix) Theory:}
A variant of heterotic M(atrix) theory is CHL M(atrix) theory~\cite{kimrey3},
which is defined as M(atrix) theory compactified on M\"obius strip, viz. 
$\Gamma_2^{\rm CHL}$ involution of heterotic M(atrix) theory compactified on 
${\bf S}_1$. As such, CHL M(atrix) theory is described by a three-dimensional 
gauge theory with eight supercharges and gauge group $\tt U(N)$ living on
dual M\"obius strip. Located at the boundary of M\"obius strip is 
(1+1)-dimensional twisted sector. 
Because $\Gamma_2^{\rm CHL}$ involution acts freely, 
the twisted sector consists only of half many $\tt N_F = 16$ chiral fermions 
than those in heterotic M(atrix) theory. 
These chiral fermions transform as a fundamental representation under the
enhanced gauge group $\tt SO(2N)$ at the boundary.

Again, the M(atrix) theory corresponding to backward boost is described by
$\tt N \rightarrow - N$ accompanied by $\tt N_F \rightarrow - N_F$ and
transpose of Young tableaux. Quite similar to heterotic M(atrix) theory, it 
is straightforward to check that the potential gauge and supersymmetry 
anomalies are cancelled completely but only if $\tt N_F \rightarrow - N_F$
as above.  To conclude, CHL M(atrix) theory defined via negative boost 
is described by $(2+1)$-dimensional gauge theory with $(0,8)$ supersymmetry
and gauge group $\tt U(N)$ living on dual M\"obius strip. At the boundary,
the gauge group is enhanced to $\tt USp(2N)$. Measured in units of fundamental 
representation of $\tt USp(2N)$, the twisted sector living at the boundary
consists of $\tt N_F = - 16$ Majorana-Weyl fermions.  

{\tt ${\bf T}_5/\integer_2$ M(atrix) Theory:}
For higher-dimensional compactification of M(atrix) theory, 
gauge theory description is not complete by itself and has to be defined
via a more fundamental fixed point quantum theory at the ultraviolet. 
Nevertheless, M(atrix) gauge theories provides useful low-energy effective
description. In particular, if the gauge theories are chiral, then absence
of gauge and supersymmetry anomalies puts severe constraint to the theory
content, hence, provides useful information. 

Consider ${\bf T}_5/\integer_2$ compactification as an example. From
D0-brane parton scattering and six-dimensional gauge and supersymmetry anomaly 
cancellation conditions, the low-energy effective description of 
corresponding M(atrix) theory has been identified with $(5+1)$-dimensional
chiral gauge theory with $(0,1)$ supersymmetry and $\tt USp(2N)$ gauge group.
The matter content consists of a hyper-multiplet from the untwisted sector 
that transforming as an anti-symmetric representation and, from the twisted sector, of thirty-two hyper-multiplets transformation as `half' 
fundamental representations.
The latter were introduced for consistency that potential gauge and 
supersymmetry anomalies are cancelled only when they are present. 
More explicitly, the gauge anomaly of untwisted sector is calculated to be
$C_4 ({\tt adjoint}) - C_4 ({\tt antisymm}) = [({\tt 2N} + 8 ) - 
({\tt 2N} - 8) ] C_4 ({\tt fundamental}) = 16 C_4 ({\tt fundamental})$.
Clearly, this is cancelled by the $\tt N_F = 32$ `half' fundamental 
representation hypermultiplets arising from the twisted sector.

Again, if the M-theory on ${\bf T}_5/\integer_2$ is boosted backward along
M-direction, according to our general consideration, the corresponding 
M(atrix) theory is defined by $\tt N \rightarrow - N$ accompanied by
$\tt N_F \rightarrow - N_F$ and weighted transpose of Young tableaux.  
The M(atrix) gauge group has become $\tt USp(-2N) = SO(2N)$. 
The negative multiplicity of twisted sector $\tt N_F = - 32$ is a 
reflection of by-now familiar fact that the background longitudinal 
five-branes are all turned into anti-five-branes.  
It is straightforward to check that the gauge anomaly cancellation 
holds satisfied even after the analytic continuation. 

We conclude that ${\bf t}^5/\integer_2$ compactification of M-theory
is described by, when boosted backward along M-direction, $(5+1)$-dimensional
chiral gauge theory with (0,1) supersymmetry and gauge group $\tt SO(2N)$. 
The twisted sector consists of $\tt N_F = - 32$ `half' fundamental 
representatoin hypermultiplets.

{\tt ${\bf T}^9/\integer_2$ M(atrix) Theory}:
Finally, consider M(atrix) theory compactified on ${\bf T}_9/\integer_2$. 
It has been shown that~\cite{kimrey4} low-energy effective description of the 
compactification is provided by (9+1)-dimensional
chiral gauge theory with (1,0) supersymmmetry and gauge group $\tt SO(32)$. 
Peculiarity of this M(atrix) gauge group is the fact that the dimension of
gauge group is completely fixed and is in fact equal to the number of 
twisted sector D0-branes $\tt N_F = 32$. This simply reflects the `holographic 
principle' in the case that all transverse directions are compactified.
Since the transverse volume is compact and finite, the would-be D0-partons
cannot be boosted indefinitely and only $\tt N_F = 32$ many D0-particles
are allowed to cancel anomalous charges from the orbifold fixed points. 

Had we boosted the M-theory compactified on ${\bf T}^9/\integer_2$ backward 
along M-direction, then from the `reciprocity relations', we expect that the 
gauge group is $\tt SO( - 32) = USp(32)$, since the twisted sector
matter multiplicity is mapped to $\tt N_F \rightarrow - N_F = - 32$. Indeed, 
$(9+1)$-dimensional $USp(n)$ super-Yang-Mills theory has gauge anomaly 
$(n + 32)$ and $n = N_F = - 32$ is the only consistent choice. 

\vskip 1.0 cm
\centerline{\bf Acknowledgments}
I thank H.~Tye and E.~Witten for invaluable discussions at early stage
of this work.


\end{document}